\documentclass[aps,prd,twocolumn,showpacs,groupedaddress]{revtex4}

\usepackage{graphicx}
\usepackage{longtable}

\begin{document}

\title{The Hamiltonian limit of (3+1)D SU(3) lattice gauge theory on anisotropic lattices}


\author{T. M. R. Byrnes}
\author{M. Loan}
\author{C. J. Hamer}
\email[]{cjh@phys.unsw.edu.au}

\affiliation{School of Physics, The University of New South Wales, Sydney, NSW 2052, Australia}

\author{Fr\'{e}d\'{e}ric D. R. Bonnet}
\altaffiliation{Department of Physics, University of Regina, Regina, SK, S4S 0A2, Canada}
\author{Derek B. Leinweber}
\author{Anthony G. Williams}
\author{James M. Zanotti}
\altaffiliation{John von Neumann-Institut f\"ur Computing NIC, Deutsches Elektronen-Synchrotron DESY, D-15738 Zeuthen, Germany}

\affiliation{Special Research Center for the Subatomic Structure of Matter (CSSM) and Department of Physics and Mathematical Physics, University of Adelaide 5005, Australia}

\date{\today}

\begin{abstract}

The extreme anisotropic limit of Euclidean SU(3) lattice gauge theory is examined to extract the Hamiltonian limit, using standard path integral Monte Carlo (PIMC) methods. We examine the mean plaquette and string tension and compare them to results obtained within the Hamiltonian framework of Kogut and Susskind. The results are a significant improvement upon previous Hamiltonian estimates, despite the extrapolation procedure necessary to extract observables. We conclude that the PIMC method is a reliable method of obtaining results for the Hamiltonian version of the theory. Our results also clearly demonstrate the universality between the Hamiltonian and Euclidean formulations of lattice gauge theory. It is particularly important to take into account the renormalization of both the anisotropy, and the Euclidean coupling $ \beta_E $, in obtaining these results. 

\end{abstract}



\maketitle

\section{Introduction}

It is well known that lattice gauge theory (LGT) can be constructed in two ways: Wilson's ``Euclidean'' \cite{wilson74} formulation where both space and time are discretized, or Kogut and Susskind's ``Hamiltonian'' \cite{kogut75} formulation where time remains continuous. The method of choice in recent years for most lattice gauge theorists has been to use classical Monte Carlo methods in the Euclidean framework to extract observables from the theory. LGT in the Hamiltonian formulation has been rather neglected in comparison. Despite this, the Hamiltonian framework still offers an interesting alternative to its Euclidean cousin. One advantage of the Hamiltonian version is that techniques familiar from quantum many-body theory can be used to attack the problem, such as strong-coupling series methods \cite{banks77}, the coupled cluster method \cite{shuohong88,bishop93,baker96,baker97,mckellar00}, the $ t $-expansion \cite{horn84,vandendoel86,morningstar92}, the plaquette expansion \cite{hollenberg93,hollenberg94,mcintosh97}, and the density matrix renormalization group (DMRG) method \cite{byrnes02}. It must be said however, these methods tend to be more successful for lower dimensional lattices. Another possible advantage is that from a numerical point of view, the reduction in the dimensionality of the lattice from four to three provides a significant reduction in computational overheads. Further, Hamiltonian results can serve as a check of the universality of Euclidean results. 

Due to the success of Monte Carlo methods in the Euclidean regime, one might expect similar levels of success in Hamiltonian LGT. Unfortunately, this has not been the case and quantum Monte Carlo methods for Hamiltonian LGT lag at least ten years behind their Euclidean counterparts. One of the first attempts at such a calculation was performed using a Green's function Monte Carlo approach by Heys and Stump for U(1) \cite{heys83,heys85} and SU(2) \cite{heys84}. Chin and co-workers soon after implemented the closely related ``guided random walk'' algorithm for U(1) in (3+1)D \cite{chin84}, then for SU(2) \cite{chin85}, and SU(3) \cite{chin86,chin88,long88}. A feature of both these methods is that a ``trial wave function'' is used to guide random walkers towards the exact wavefunction. This appeared in the early days to be an advantage of the technique as the physical features of the wavefunction may be put in by hand, while the unknown part may be found through the stochastic process. However, later investigations \cite{hamer00} have shown that there is an unacceptable dependence of the observables on the parameters of the wavefunction, in the case of SU(3) theory in (3+1)D. Other methods, such as the ``projector Monte Carlo'' method \cite{blankenbecler83,degrand85} and the related ``stochastic truncation'' method \cite{allton89} have a version of the ``minus-sign problem'' arising due to the strong coupling (electric field) representation used for the basis states. This is due to the necessary introduction of Clebsch-Gordan coefficients for non-Abelian theories, which potentially cause destructive interference between the transition amplitudes. In view of the lack of any clear success of these quantum Monte Carlo methods, we are therefore forced to pursue an alternative approach. 

In a previous study, standard Euclidean path integral Monte Carlo (PIMC) methods were used to extract the Hamiltonian limit for the U(1) lattice gauge theory in (2+1) dimensions \cite{loan02}. The basic idea is to measure observables on increasingly anisotropic lattices, then extrapolate to the Hamiltonian limit, corresponding to $ \Delta \tau = a_t/ a_s \rightarrow 0 $, where $ a_t $ and $ a_s $ are the lattice spacings in the time and space directions respectively. The results obtained \cite{loan02} show excellent agreement between the extrapolated results and Hamiltonian estimates of the same quantities. In this work we attempt a similar procedure for pure SU(3) gauge theory in (3+1) dimensions. We calculate two basic quantities, the average plaquette and the string tension, and compare them to results obtained in the Hamiltonian formulation. We find that indeed the anisotropic Euclidean results converge to the Hamiltonian estimates, once the difference of scales \cite{hasenfratz81} between the two theories has been taken into account. Specifically we find that this is particularly important at {\it finite} anisotropy \cite{karsch82} in the extrapolation procedure. This will be discussed in more detail in section \ref{sec:su3}.

The use of anisotropic lattices has generated much interest in recent years, following the work of Morningstar and Peardon \cite{morningstar97,morningstar99} in extracting accurate glueball masses. This same approach has now been extended to extracting heavy quark spectra \cite{chen01,okamoto02}. The importance of taking into account the difference of scales on anisotropic lattices in these studies has already been discussed by Klassen \cite{klassen98}, through the renormalization of the anisotropy. In our case, the renormalization of the the isotropic Euclidean coupling $ \beta_E $ is also important in extracting our results. This may be of some relevance to other studies on anisotropic lattices.

In Sec. \ref{sec:su3} we briefly discuss the SU(3) model as defined on anisotropic lattices. We also discuss in this section our extrapolation procedure to the Hamiltonian limit. In Sec. \ref{sec:method} we explain the simulation methods used to obtain our results, and in Sec. \ref{sec:results} we show our results for the average plaquette and string tension. Some concluding remarks are given in Sec. \ref{sec:conc}.

\section{The anisotropic SU(3) model}
\label{sec:su3}

The anisotropic action is given by \cite{creutz83}
\begin{equation}
\label{ani_action_1}
S = \frac{\beta_\sigma}{\xi} \sum_x \sum_{i>j ; \hspace{1mm} i \ne 4} \left( 1 - P_{ij}(x) \right)  + \beta_\tau \xi \sum_x \sum_{i \ne 4} \left( 1 - P_{4i}(x) \right) ,
\end{equation}
where the first term sums over all space-like plaquettes on the lattice, and the second term sums over time-like plaquettes. A plaquette at lattice position $ x $ is defined by
\begin{equation}
\label{plaquettedefinition}
P_{\mu \nu}(x) = \frac{1}{3} \mbox{Re}\mbox{Tr} \left[ U_\mu (x) U_\nu (x+ \hat{\mu}) U_\mu^\dagger (x + \hat{\nu}) U_\nu^\dagger (x) \right] ,
\end{equation} 
where $ U_\mu (x) $ is the SU(3) gauge field variable. The couplings $ \beta_\sigma $ and $ \beta_\tau $ are defined by 
\begin{equation}
\label{beta_def}
\beta_\sigma = \frac{6}{g_\sigma^2}, \hspace{1cm} \beta_\tau = \frac{6}{g_\tau^2},
\end{equation}
and the anisotropy factor, or aspect ratio is defined by 
\begin{equation}
\xi = \frac{1}{\Delta \tau} = \frac{a_s}{a_t}.
\end{equation}
One must include different space-like and time-like couplings $ g_\sigma $ and $ g_\tau $ in Eq. (\ref{beta_def}) in order to allow the freedom to renormalize so that correlation lengths are equal in both directions, even though the spacings $ a_s $ and $ a_t $ are different. In the continuum limit we require that physical quantities be independent of changes in both of these quantities. Therefore we require the two couplings to be a function of both $ a_s $ and $ a_t $. Karsch \cite{karsch82}, using the background field method of Dashen and Gross \cite{dashen81} and Hasenfratz and Hasenfratz \cite{hasenfratz81}, obtained a mapping between the equivalent couplings of the Euclidean and anisotropic actions
\begin{eqnarray}
\label{karsch_g_sigma}
\frac{1}{g_\sigma^2 } & = & \frac{1}{g_E^2} + c_\sigma (\xi) + O(g_E^2), \\
\label{karsch_g_tau}
\frac{1}{g_\tau^2} & = & \frac{1}{g_E^2} + c_\tau (\xi) + O(g_E^2) ,
\end{eqnarray}
where $ g_E $ is the coupling for the Euclidean theory with $ \xi = 1 $. The factors $ c_\sigma (\xi) $ and $ c_\tau (\xi) $ are defined in Eqs. (2.24) and (2.25) of Ref. \cite{karsch82}. For the Euclidean theory with $ \xi = 1 $, these factors both approach zero, and therefore we recover the usual action with only one coupling $ g_\sigma = g_\tau = g_E $. In the limit $ \xi \rightarrow \infty $, Eqs. (\ref{karsch_g_sigma}) and (\ref{karsch_g_tau}) reduce to the Hamiltonian values as obtained by Hasenfratz and Hasenfratz \cite{hasenfratz81}. 

It is convenient to rewrite the action (\ref{ani_action_1}) in a more symmetric fashion
\begin{equation}
\label{ani_action_2}
S = \frac{\beta_\xi}{\bar{\xi} } \sum_x \sum_{i>j ; \hspace{1mm} i \ne 4} \left( 1 - P_{ij}(x) \right) + \beta_\xi \bar{\xi} \sum_x \sum_{i \ne 4} \left( 1 -  P_{4i}(x) \right),
\end{equation}
where
\begin{equation}
\beta_\xi = \frac{6}{g_\xi^2}, \hspace{1cm} \bar{\xi} = \frac{\xi}{\eta},
\end{equation}
with $ g_\xi^2 = g_\sigma g_\tau $ and $ \eta = ( g_\tau^2 / g_\sigma^2)^{1/2} $. In the limit $ \xi \rightarrow \infty $, $ \beta_\xi $ goes to the Hamiltonian coupling $ \beta_H $. Using Eqs. (\ref{karsch_g_sigma}) and (\ref{karsch_g_tau}) we may write a correspondence between these variables and the Euclidean theory
\begin{eqnarray}
\label{karsch_betaxi}
\beta_\xi & = & \beta_E + 3(c_\sigma (\xi) + c_\tau (\xi)) + O(\beta_E^{-1}) \\
\label{karsch_eta}
\eta & = & 1 + \frac{3}{\beta_E}(c_\sigma(\xi) - c_\tau (\xi)) + O(\beta_E^{-2}) ,
\end{eqnarray}
where $ \beta_E = 6/g_E^2 $. Therefore for every $ (\beta_E, \xi) $ pair there is a corresponding pair of couplings $ (\beta_\xi, \bar{\xi}) $. The relation between $ \xi $ and $ \bar{\xi} $ has been discussed in some detail by Klassen \cite{klassen98}. In his language, $ \bar{\xi} $ is the bare anisotropy, while $ \xi $ is the renormalized anisotropy. To evaluate the factors $ c_\sigma( \xi) $ and $ c_\tau (\xi) $, one may either directly calculate them in terms of the integrals given in Ref. \cite{karsch82}, or in the case of Eq. (\ref{karsch_eta}), use the parameterization given by Klassen \cite{klassen98}.

\begin{figure}
\scalebox{0.45}{\includegraphics{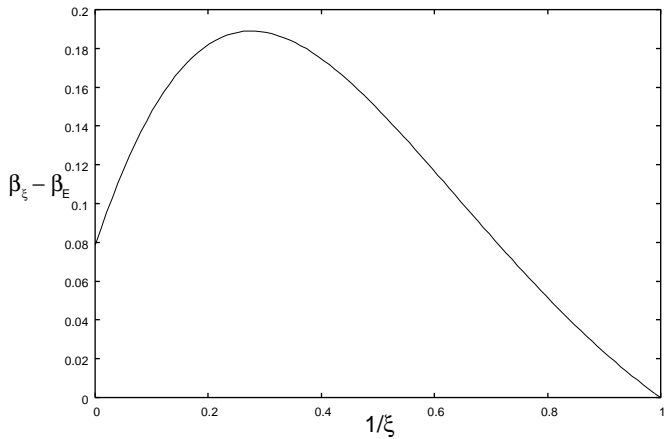}}
\caption{
\label{figkarsch}
The discrepancy between the anisotropic and Euclidean couplings as a function of the anisotropy.
}
\end{figure}

We now discuss our extrapolation procedure in order to obtain Hamiltonian estimates from an anisotropic lattice. In a naive extrapolation procedure, one might assume $ \beta = \beta_\sigma = \beta_\tau $ in Eq. (\ref{ani_action_1}), and extrapolate physical quantities at constant $ \beta $ to the Hamiltonian limit, $ \xi \rightarrow \infty $. This procedure is incorrect, however, because $ \beta_\sigma \neq \beta_\tau \neq \beta $ due to renormalization. The correct procedure is to extrapolate to $ \xi \rightarrow \infty $ at constant $ \beta_E $. We may summarize our procedure as follows:
\begin{enumerate}
\item For a particular $ \beta_E $, choose several anisotropies $ \xi $. 
\item Calculate the corresponding values of $ \beta_\xi $ and $ \bar{\xi} $ using Eqs. (\ref{karsch_betaxi}) and (\ref{karsch_eta}).
\item Use the action given in (\ref{ani_action_2}) to calculate physical observables at these couplings. 
\item Perform a polynomial fit to the data in the inverse square anisotropy 
$ \Delta \tau^2 $ at constant $ \beta_E $, and extrapolate measured quantities 
to $ \xi \rightarrow \infty $ ($\Delta \tau \rightarrow 0 $) (we extrapolate in
 the inverse square anisotropy since the leading discretization errors in the 
timelike direction are expected to be of order $ a_t^2 $.). 
\end{enumerate}
It is interesting to note that the discrepancy between $ \beta_\xi $ and $ \beta_E $ in Eq. (\ref{karsch_betaxi}) reaches a maximum around $ \xi \approx 3.6215 $, as can be seen from Fig. \ref{figkarsch}. We see that in the vicinity of the maximum, the discrepancy reaches nearly $ \beta_\xi - \beta_E \approx 0.19 $. Since the most anisotropic lattice we use is in the vicinity of this maximum, significant discrepancies begin to creep in if the correct procedure is not used.

\section{Method}

\label{sec:method}

\subsection{Simulation Details}

\label{method:sim}

We analyze the action given in Eq. (\ref{ani_action_2}) by standard path integral Monte Carlo methods. Configurations are generated using the Cabibbo-Marinari \cite{cabibbo82} pseudo-heat bath algorithm, applied to the three diagonal SU(2) subgroups of SU(3). To analyze the behavior between weak and strong coupling we concentrate on the region $ \beta_\xi = 4 - 7 $. For each $ \beta_\xi $ value we generate configurations for anisotropies in the range $ \bar{\xi} = 1 - 3 $. These parameters $ (\beta_\xi, \bar{\xi}) $ may be converted into their corresponding $ (\beta_E,\xi ) $ by solving (\ref{karsch_betaxi}) and (\ref{karsch_eta}) numerically. Our full set of parameters is shown in Table \ref{tab:parameters}, together with the equivalent $ (\beta_E, \xi) $ values.

\begin{table}
\caption{
\label{tab:parameters}
Parameters used for each configuration set. 
}
\begin{ruledtabular}
\begin{tabular}{ccccc}
Dimensions & $ \beta_\xi $ & $ \bar{\xi} $ & $ \beta_E $ & $ \xi $ \\
\hline
$ 8^3 \times 8 $ & 4.0 & 1 & 4.0 & 1.0 \\
$ 8^3 \times 8 $ & 5.0 & 1 & 5.0 & 1.0 \\
$ 8^3 \times 8 $ & 5.4 & 1 & 5.4 & 1.0 \\
$ 8^3 \times 8 $ & 5.6 & 1 & 5.6 & 1.0 \\
$ 8^3 \times 8 $ & 5.8 & 1 & 5.8 & 1.0 \\
$ 8^3 \times 8 $ & 6.0 & 1 & 6.0 & 1.0 \\
$ 8^3 \times 8 $ & 7.0 & 1 & 7.0 & 1.0 \\
$ 8^3 \times 12 $ & 4.0 & $\sqrt{3/2}$ & 3.9396 & 1.2968 \\
$ 8^3 \times 12 $ & 5.0 & $\sqrt{3/2}$ & 4.9428 & 1.2795 \\
$ 8^3 \times 12 $ & 5.4 & $\sqrt{3/2}$ & 5.3437  & 1.2747 \\
$ 8^3 \times 16 $ & 4.0 & $ \sqrt{2}$ & 3.8992 & 1.5444  \\
$ 8^3 \times 16 $ & 5.0 & $ \sqrt{2}$ & 4.9036 &  1.5139 \\
$ 8^3 \times 16 $ & 5.4 & $ \sqrt{2}$ & 5.3049  & 1.5053 \\
$ 8^3 \times 12 $ & 5.0 & 3/2 & 4.8891 & 1.6197 \\
$ 8^3 \times 12 $ & 5.4 & 3/2 & 5.2904 & 1.6095 \\
$ 8^3 \times 12 $ & 5.6 & 3/2 & 5.4910 & 1.6050 \\
$ 8^3 \times 12 $ & 5.8 & 3/2 & 5.6916 & 1.6008 \\
$ 8^3 \times 12 $ & 6.0 & 3/2 & 5.8921 & 1.5970 \\
$ 8^3 \times 12 $ & 7.0 & 3/2 & 6.8941 & 1.5815 \\
$ 8^3 \times 24 $ & 4.0 & $ \sqrt{3}$ & 3.8546 &  1.9569 \\
$ 8^3 \times 24 $ & 5.0 & $ \sqrt{3}$ & 4.8588 &  1.9053  \\
$ 8^3 \times 24 $ & 5.4 & $ \sqrt{3}$ & 5.2602 &  1.8907 \\
$ 8^3 \times 16 $ & 5.0 & 2 & 4.8366 & 2.2341 \\
$ 8^3 \times 16 $ & 5.4 & 2 & 5.2376 & 2.2146 \\
$ 8^3 \times 16 $ & 5.6 & 2 & 5.4381 & 2.2060 \\
$ 8^3 \times 16 $ & 5.8 & 2 & 5.6385 & 2.1981 \\
$ 8^3 \times 16 $ & 6.0 & 2 & 5.8389 & 2.1908 \\
$ 8^3 \times 16 $ & 7.0 & 2 & 6.8406 & 2.1609 \\
$ 8^3 \times 24 $ & 4.0 & 3 & 3.8110 & 3.5811 \\
$ 8^3 \times 24 $ & 5.0 & 3 & 4.8112 & 3.4538 \\
$ 8^3 \times 24 $ & 5.4 & 3 & 5.2113 & 3.4171 \\
$ 8^3 \times 24 $ & 5.6 & 3 & 5.4114 & 3.4010 \\
$ 8^3 \times 24 $ & 5.8 & 3 & 5.6114 & 3.3860 \\
$ 8^3 \times 24 $ & 6.0 & 3 & 5.8115 & 3.3721 \\
$ 8^3 \times 24 $ & 7.0 & 3 & 6.8117 & 3.3152 \\
\end{tabular}
\end{ruledtabular}
\end{table}

\begin{figure}
\scalebox{0.45}{\includegraphics{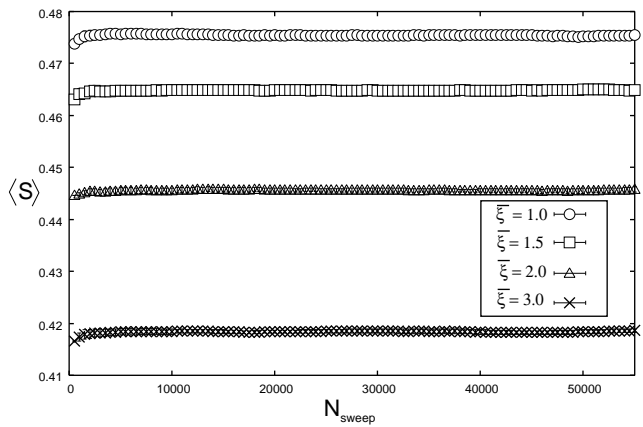}}
\caption{
\label{figequilibration}
Evolution of the average action $ \langle S \rangle $ with the number of sweeps of the lattice, for $ \beta_\xi = 5.6 $, and various anisotropies. 
}
\end{figure}

Due to the large number of configuration sets that must be generated, we limit 
ourselves to relatively modest lattice sizes, ranging from $ 8^3 \times 8 $ to 
$ 8^3 \times 24 $. We adjust the lattice size in the time direction according 
to the anisotropy used in order to keep the physical length in the time and 
space directions equal. Configurations are given a cold start, and then 5000 
thermalization sweeps in order to equilibrate. After thermalization, 
configurations are stored every 500 sweeps, for 100 configurations. As a 
measure of the equilibration, we plot the average action for various 
anisotropies in Fig. \ref{figequilibration}. We see that for each value of the 
anisotropy the configurations relax to equilibrium after of order $ 1000 $ 
sweeps. In particular we see very little difference in equilibration times for 
the various anisotropies. This is in contrast to the results of Ref. 
\cite{loan02}, which used a standard Metropolis algorithm combined with
Fourier acceleration techniques for the U(1) model in (2+1) dimensions.
There it was found that equilibration times were longer for the anisotropic 
lattices, 
despite using Fourier acceleration techniques. 
The advantage of the pseudo-heat bath algorithm in updating
anisotropic lattices, where the space-time $ \beta_\tau $ and  $\xi$
can be very large, is most easily seen when considering its
application to SU(2)-color gauge theory.  In this case a link
variable is updated without any reference to the current link.
Rather only the staples associated with the link are considered
in making the SU(2) update.  The improved ergodicity afforded
by this algorithm is largely carried over to the SU(2)
subgroup updates of SU(3) links.

According to the extrapolation procedure described in Sec. \ref{sec:su3}, we must extrapolate at constant $ \beta_E $, not $ \beta_\xi $ as we have calculated. Therefore for all observables we must interpolate our values in the $ (\beta_\xi, \bar{\xi}) $ plane onto lines of constant $ \beta_E $. We do this using the modified Shepard method \cite{renka88} of multivariate interpolation. We choose our interpolation points along the lines $ \beta_E = 3.8110, 4.8112, 5.2113, 5.4114, 5.6114, 5.8115, 6.8117 $, and $ \xi $ values such that they are as close as possible to existing data points. These points are chosen such that there is no interpolation for the most anisotropic points ($ \bar{\xi} = 3 $), and there is minimal interpolation for the remaining points, to minimize the interpolation error. In practice these interpolation errors are quite negligible compared with the original errors on the data points, such that the error on the interpolated value can be taken as the same as the error in the nearest data point.

Clearly a more straightforward procedure would have been to generate data points originally such that they were constant in $ \beta_E $, by choosing appropriate varying $ (\beta_\xi, \bar{\xi} ) $ values. In a future study we would adopt such a procedure, although there is little loss in accuracy for our current results.

\subsection{The Static Quark Potential}

\begin{figure}
\scalebox{0.45}{\includegraphics{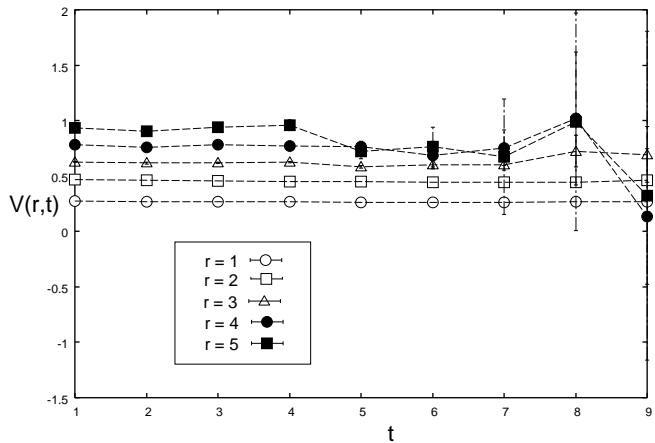}}
\caption{
\label{figplateau}
Effective potential plots for $ \beta_\xi = 5.60 $ and $ \bar{\xi} = 2.0 $. 
}
\end{figure}

\begin{figure}
\scalebox{0.45}{\includegraphics{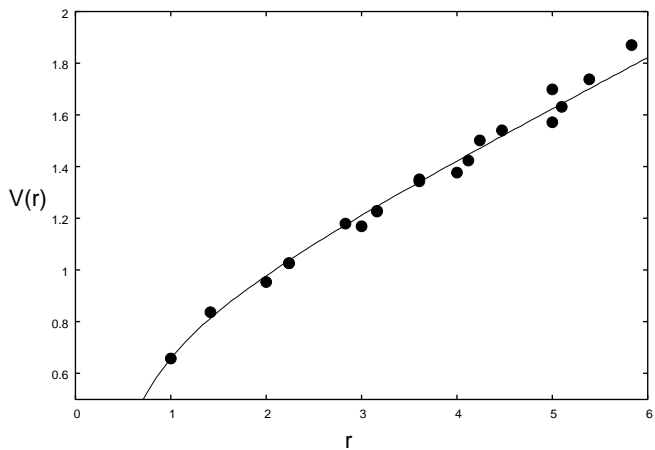}}
\caption{
\label{figpotential}
Static quark potential at $ \beta_\xi = 5.8 $ and $ \bar{\xi} = 3.0 $. 
}
\end{figure}

The static quark potential is extracted from the expectation values of Wilson loops, which are expected to behave like
\begin{equation}
W(r,t) = \sum_i C_i (r) \exp(-V_i (r) t), 
\end{equation}
where the summation is over the excited state contributions to the expectation value, and $ i=1 $ corresponds to the contribution from the ground state. To obtain the optimal signal-to-noise ratio, we must suppress the contributions from the excited states, which may be done by APE smearing \cite{falcioni85,albanese87}. This involves replacing a particular space-like link by
\begin{eqnarray}
\nonumber
U_\mu (x) & \rightarrow & P \Big[ (1-\alpha) U_\mu (x) \\
\nonumber
& & + \frac{\alpha}{4} \sum_{\nu=1 ;  \nu \ne \mu}^3 \big[ U_\nu (x) U_\mu (x+ \hat{\nu}) U^\dagger_\nu (x + \hat{\mu}) \\
\nonumber
& & + U_\nu^\dagger (x - \hat{\nu}) U_\mu(x-\hat{\nu}) U_\nu (x - \hat{\nu} + \hat{\mu}) \big] \Big], \\
\end{eqnarray}
where the $ P $ denotes a projection back onto SU(3), and $ \alpha $ is a parameter which is defined by the user. This is repeated $ n_{\mbox{\footnotesize{APE}}} $ times. This amounts to adding in a fraction $ \alpha $ of neighboring staples for every space-like link on the lattice. To tune the smearing parameters, we fix the smearing fraction to $ \alpha = 0.7 $, as it is sufficient to fix $ \alpha $ and tune $ n_{\mbox{\footnotesize{APE}}} $ \cite{bonnet00}. To find the optimum $ n_{\mbox{\footnotesize{APE}}} $, we look for the cleanest plateaus in the effective potential
\begin{equation}
V(r,t) = \ln \left[ \frac{W(r,t)}{W(r,t+1)} \right],
\end{equation}
and also examine the ratio \cite{bonnet99}
\begin{equation}
W^{t+1}(r,t)/W^{t} (r,t+1), 
\end{equation}
which should be near unity for good ground state dominance. A typical
value which proved to be sufficient for most cases was $ n_{\mbox{\footnotesize{APE}}} =5 $,
although for small $ \beta_\xi $ the overall effectiveness of the
smearing procedure was reduced. An example of a typical effective potential plot is shown in Fig. \ref{figplateau}, for $ \beta_\xi = 5.60 $ and $ \bar{\xi} = 2.0 $. We see good plateau behavior for small $ t $ values, which reflects the optimum smearing. Finally we extract the string tension by fitting with the form
\begin{equation}
\label{potentialfit}
V(r) = V_0 + K r - e/r ,
\end{equation}
where $ e = \pi/12 $ \cite{luscher81},  and $ K $ and $ V_0 $ are fit
parameters.  The string tension is then found according to $ K = \sigma
a^2 $.   
An example of such a fit is shown in Fig. \ref{figpotential}, for $ \beta_\xi = 5.8 $ and $ \bar{\xi} = 3.0 $. We see that the data are fitted very well by (\ref{potentialfit}), in this instance giving $ K = \sigma a^2  = 0.189(4) $ and $ V_0 = 0.732(7) $. 

\section{Results}
\label{sec:results}

\subsection{Average Plaquette}
\label{sec:results.avplaq}

\begin{figure}
\scalebox{0.45}{\includegraphics{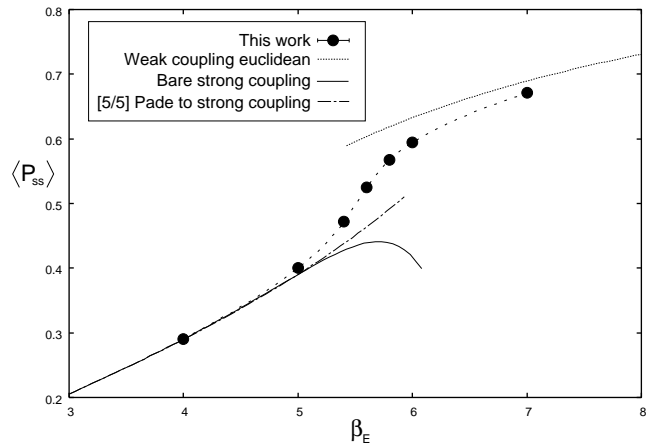}}
\caption{
\label{fig_avplaq_eucl}
Comparison of our PIMC data for the average plaquette with strong and weak coupling expansions for the isotropic lattice $ \xi = 1 $. Dashed lines are merely to guide the eye.
}
\end{figure}

Our aim in this section is to obtain estimates for the average plaquette,
defined as the expectation value of Eq. (\ref{plaquettedefinition}), for 
the Hamiltonian theory. As no time-like plaquettes are present in the 
Hamiltonian lattice, we therefore only calculate space-like plaquettes in 
the Euclidean theory. Henceforth we will refer to the average space-like 
plaquette simply as ``average plaquette'', unless specified otherwise. As 
a first test of our results, we compare the average plaquette for the 
isotropic lattice $ \xi = 1 $ to existing results. This is shown in Fig. 
\ref{fig_avplaq_eucl}, together with strong coupling expansions to order 
$ \beta_E^{15}  $ \cite{balian75,smit82}, and weak coupling expansions to 
order $ \beta_E^{-2} $ \cite{digiacomo81}. We also show a [5/5] Pad{\'e} 
extrapolation of the strong coupling series, which shows that beyond $ 
\beta_E \approx 5 $ the series diverges away. Our PIMC data match 
smoothly onto the strong coupling and weak coupling expansions in their 
respective limits, as expected \footnote{We note that the physical time 
extent of the $ \beta_E=7.0 $ lattices is sufficiently small to take us 
beyond the deconfinement phase transition. However, the small spatial 
lattice extent constrains us to examine short distance quantites which 
are not severely affected by deconfinement issues. Therefore, we present 
results for the mean plaquette at $ \beta_E =7.0 $ with this 
reservation.}.  
It can be seen that the crossover from 
strong coupling to weak coupling behavior takes place in the region $ \beta_E \approx 5-6 $, as is well known. 

\begin{figure}
\scalebox{0.45}{\includegraphics{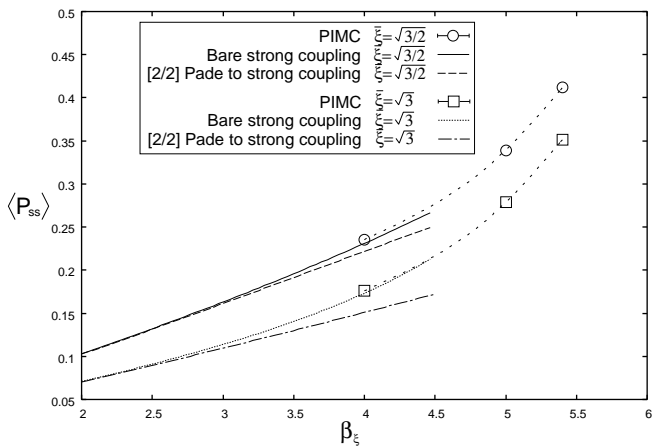}}
\caption{
\label{fig_ani_strongcoup}
Comparison of the average space-like plaquette calculated using PIMC at anisotropies $ \bar{\xi} = \sqrt{3/2}$ and $ \bar{\xi} = \sqrt{3} $ with strong coupling expansions to order $ \beta_\xi^6 $. Dashed lines are to guide the eye.
}
\end{figure}

At finite anisotropy, we are not aware of any results available to make a similar comparison. It is fairly straightforward however to calculate the corresponding strong coupling expansion to order $ \beta_\xi^6 $. The calculations for this are given in the Appendix. Setting $ I = J = 1 $ into (\ref{Wssani}) gives
\begin{eqnarray}
\nonumber
\langle P_{ss} \rangle & = & \frac{1}{3 \bar{\xi} } b + \frac{1 }{6 \bar{\xi}^2} b^2 - \frac{5 }{ 72 \bar{\xi}^4 } b^4  + \left( \frac{2 \bar{\xi}^3 }{243} - \frac{65}{1944  \bar{\xi}^5} \right) b^5 \\
\label{strong_coup_ani}
& & + \left( \frac{\bar{\xi}^2}{81 } + \frac{4 \bar{\xi}^4}{243} + \frac{749}{19440 \bar{\xi}^6} \right) b^6  + O( b^7) ,
\end{eqnarray}
where $ b = \beta_\xi/6 $. A comparison between the strong coupling expansion and our PIMC data is shown in Fig. \ref{fig_ani_strongcoup}, along with a [2/2] Pad{\'e} extrapolation. We see consistent results between the expansions and our data for $ \beta_\xi =  4 $, which is still in the strong coupling region. There is a trend for the series with larger values of the anisotropy $ \bar{\xi} $ to break down earlier as can be seen from the Pad{\'e} approximant for $ \bar{\xi} = \sqrt{3} $.

\begin{figure}
\scalebox{0.45}{\includegraphics{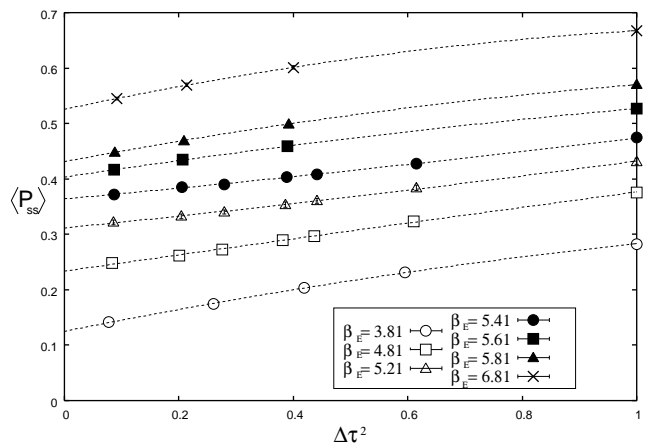}}
\caption{
\label{fig_avplaq_extrap}
Extrapolation of the average plaquette to the Hamiltonian limit $ \Delta \tau \rightarrow 0 $, for fixed $ \beta_E $. Dashed lines show quadratic fits to the data in $ \Delta \tau^2 $. }
\end{figure}

We now extrapolate our results to the Hamiltonian limit. Performing the 
interpolation as described in Sec. \ref{method:sim}, we obtain points at constant $ \beta_E $. Our results are then extrapolated to the Hamiltonian limit in powers of $ \Delta \tau^2 $, as shown in Fig. \ref{fig_avplaq_extrap}
 We see a fairly smooth dependence on 
$ \Delta \tau^2 $, for all $ \beta_E $. Error estimates for the extrapolation 
may be estimated by comparing fits of different orders (linear, quadratic, 
cubic, ...) in $ \Delta \tau^2 $. Our numerical values in the 
$ \Delta \tau \rightarrow 0 $ limit are shown in Table \ref{tab:results}.

\begin{table}
\caption{
\label{tab:results}
A summary of our estimates in the Hamiltonian limit $ \Delta \tau \rightarrow 0 $ for the average (space-like) plaquette and the string tension. The Hamiltonian coupling $ \lambda $, calculated from Eq. (\ref{euchamrelation}), is also shown.
}
\begin{ruledtabular}
\begin{tabular}{cccc}
$ \beta_E $ & $ \lambda $ & $ \langle P_{ss} \rangle  $ & $ \sigma a^2 $ \\
\hline
3.8110 & 2.5214 &    0.125(5)  & \\
4.8112 & 3.9849 &    0.23(1)   & 1.08(20) \\
5.2113 & 4.6637 &    0.312(7)  & 0.57(4)  \\
5.4114 & 5.0231 &    0.36(1)   & 0.31(2)  \\
5.6114 & 5.3958 &    0.403(5)  &0.17(1)   \\
5.8115 & 5.7819 &    0.431(6)  & 0.110(4)\\
6.8117 & 7.9125 &    0.526(7)   & \\
\end{tabular}
\end{ruledtabular}
\end{table}

\begin{figure}
\scalebox{0.45}{\includegraphics{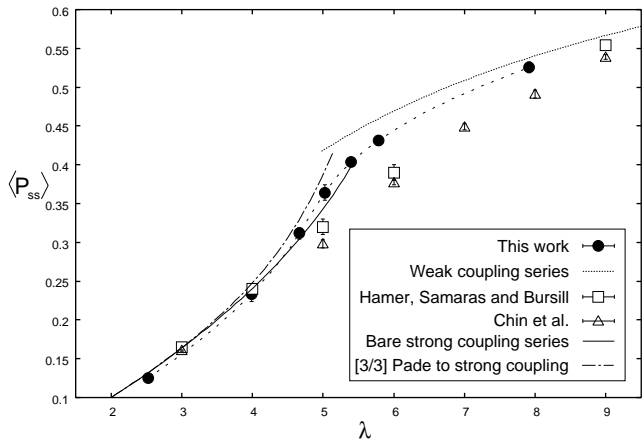}}
\caption{
\label{fig_plaq_ham}
Our extrapolated Euclidean PIMC results compared to Hamiltonian strong coupling \cite{hamer86}, Hamiltonian weak coupling \cite{hofsass83}, and Green's Function Monte Carlo (GFMC) results \cite{hamer00}. Dashed lines are merely to guide the eye.
}
\end{figure}

To compare our extrapolated results to existing Hamiltonian results, we must take into account the difference of scales between the two regimes. The Hamiltonian coupling parameter $ \lambda = 6/g_H^4 $, where $ g_H \equiv \lim_{\xi \rightarrow \infty} g_\xi $, may be related to the Euclidean coupling through the relation \cite{hasenfratz81}
\begin{equation}
\label{euchamrelation}
\beta_E = \sqrt{6 \lambda} - 0.07848. 
\end{equation}
Using this relation we may plot our extrapolated Euclidean estimates against previous results obtained within the Hamiltonian formulation, as shown in Fig. \ref{fig_plaq_ham}. This figure compares our present results with Green's Function Monte Carlo (GFMC) results of Chin {\it et al.} \cite{chin88} and Hamer, Samaras, and Bursill \cite{hamer00}, the weak coupling series to order $ \lambda^{-1/2} $ of Hofs{\"a}ss and Horsley \cite{hofsass83}, and the strong coupling series to order $ \lambda^7 $ of Hamer, Irving and Preece \cite{hamer86}. For the strong coupling series we show both the raw series sum, and the [3/3] Pad{\'e} approximant. The analysis of Hamer {\it et al.} \cite{hamer00} shows that finite-size effects in the Monte Carlo data should be negligible at this level of accuracy. 

For $ \lambda < 4 $, the agreement between the different estimates is excellent, except that our point at $ \lambda = 2.52 $ is a little lower than the strong coupling series result. A possible explanation is that we have relied heavily on the weak-coupling relations between scales of the anisotropic and Hamiltonian theories and the Euclidean theory (i.e. Eqs. (\ref{karsch_betaxi}), (\ref{karsch_eta}), and (\ref{euchamrelation})). It is possible higher order terms in these expansions begin to contribute larger effects at $ \lambda < 3 $. 

In the region $ \lambda = 5-8 $, however, there is a large discrepancy between our results and the previous GFMC estimates \cite{chin88,hamer00}. We find that the crossover from strong to weak coupling behavior occurs at much the same couplings as in the isotropic regime, and earlier than shown by the GFMC results. There is quite good agreement, on the other hand between our results and the strong coupling series extrapolations in this regime. This gives us confidence that our present results are the more accurate. It was pointed out in Ref. \cite{hamer00} that the GFMC method suffers from an unacceptable dependence on the trial wave function, and we surmise that the variational parameter may not have been optimized to the correct value in this region. 

Further evidence is provided by Long {\it et al.} \cite{long88} who showed that GFMC results for the ground-state energy change significantly if a second variational parameter is added in to the trial wave function. We conclude that the PIMC offers a more unbiased method of extracting results, even with the somewhat undesirable extra step of having to perform a $ \Delta \tau \rightarrow 0 $ extrapolation.

\subsection{String Tension}
\label{sec:results.strtension}

\begin{figure}
\scalebox{0.45}{\includegraphics{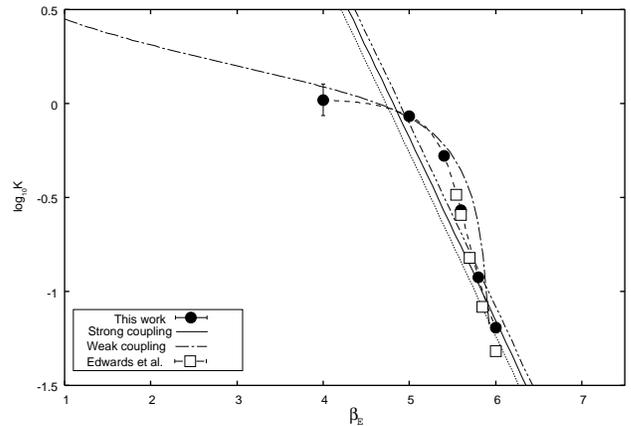}}
\caption{
\label{fig_eucl_scaling}
The string tension at $ \xi = 1 $. We plot our own PIMC results, together with a 12th order strong coupling expansion \cite{munster80}. }
\end{figure}

We now turn to calculating the string tension. We again use only 
space-like Wilson loops, so we can ultimately compare to Hamiltonian 
results. Fig. \ref{fig_eucl_scaling} shows results for the isotropic case, $ \xi = 1 
$. Our results for small $ \beta_E $ are contaminated by large errors due 
to the fast exponential decay of the Wilson loops. 
The parallel lines show the two-loop scaling form
\begin{equation}
\label{scalingeqn}
\Lambda = \frac{1}{a} (\gamma_0 g^2)^{-\gamma_1/2 \gamma_0^2} \exp 
\left( -\frac{1}{2 \gamma_0 g^2} \right),
\end{equation}
where
\begin{equation}
\gamma_0 = \frac{11}{16 \pi^2}, \hspace{1cm} \gamma_1 = \frac{102}{(16 
\pi^2)^2},
\end{equation}
The coupling $ g $ refers to the Euclidean coupling $ g_E $ in this 
case, and the scaling parameter here is $ \Lambda = \Lambda_E $, the Euclidean 
scaling parameter. Our data asymptotically appear to approach the 
expected scaling form 
in the weak coupling region. The parallel lines
correspond to
\begin{equation}
\sqrt{\sigma} = (113 \pm 10)\Lambda_E ,
\end{equation}
or
\begin{equation}
\sqrt{\sigma} = (1.35 \pm 0.12 ) \Lambda_{\mbox{\footnotesize mom}} ,
\label{eq21}
\end{equation}
where $ \Lambda_{\mbox{\footnotesize mom}} $ is the perturbative QCD scale 
parameter \cite{hasenfratz80,hasenfratz81}. 
It is well known, however, and is evident from Fig. 
\ref{fig_eucl_scaling}, that in quantitative terms the string tension
follows neither two-loop or three-loop scaling at these couplings; and a
more sophisticated fit by Edwards {\it et. al.} \cite{edwards98} taking into account
various correction terms gives a much smaller
value,
\begin{equation}
\sqrt{\sigma} = (75 \pm 1)\Lambda_E ,
\label{eq22}
\end{equation}
We do not attempt such a fit here.

On the same figure are plotted the strong coupling expansions of M{\"u}nster and Weisz to 12th order \cite{munster80},
and the Monte Carlo results of Edwards { \it et al.} \cite{edwards98} . Due to 
the presence of the roughening transition at $ \beta_E \approx 5.5 $, we expect
 the strong coupling expansion to diverge from the data beyond this point, as 
is seen in the figure. 
At the smaller values of $\beta_E$, our results agree well with those of
Edwards { \it et al.} \cite{edwards98}; but at larger $\beta_E$, our results are
higher than theirs. This is most probably due to the smaller lattice
size used in our calculations, as discussed in Ref \cite{edwards98}, and is 
consistent with the higher
estimate of the string tension obtained at equation (\ref{eq21}). One
would need to use larger lattice sizes or else an improved action to get
more accurate estimates of the asymptotic parameters.

\begin{figure}
\scalebox{0.45}{\includegraphics{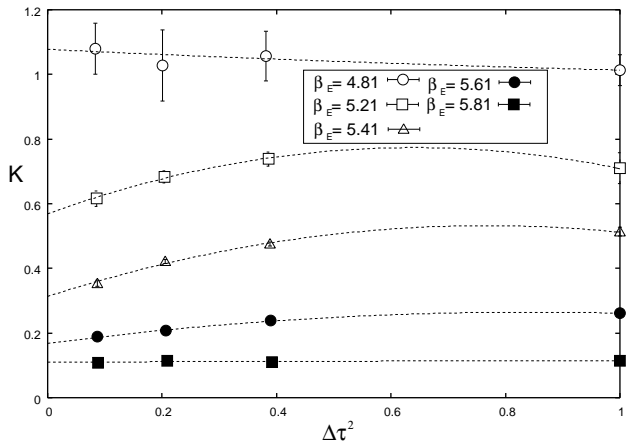}}
\caption{
\label{fig_stringtens_extrap}
Extrapolation of the string tension to the Hamiltonian limit $ \Delta \tau \rightarrow 0 $. Dashed lines show quadratic fits in $ \Delta \tau^2 $. 
}
\end{figure}

The extrapolation to the Hamiltonian limit is shown in Fig. \ref{fig_stringtens_extrap}. This is performed again in powers of $ \Delta \tau^2 $, in a similar fashion to the average plaquette. There is a trend towards a stronger curvature in the extrapolation for smaller values of $ \beta_E $, suggesting that our estimate for $ \beta_E = 4.81 $ is probably too high. There are however large error bars on these data points.

\begin{figure}
\scalebox{0.45}{\includegraphics{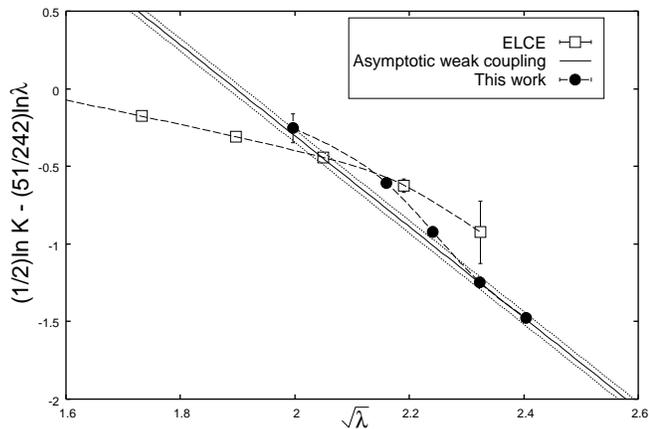}}
\caption{
\label{fig_stringtens_ham}
Extrapolated PIMC results together with ELCE results, and GFMC estimates of the string tension. 
}
\end{figure}

In Fig. \ref{fig_stringtens_ham}, our extrapolated results for the string tension are 
compared with Hamiltonian estimates obtained by Hamer, Irving and Preece 
\cite{hamer86} using an Exact Linked Cluster Expansion (ELCE) method. 
Our axes are such that the scaling relation (\ref{scalingeqn}) for the 
string tension appears as a straight line in the plot. In this case we use 
the Hamiltonian versions of the couplings, $ g = g_H $, $ \Lambda = 
\Lambda_E $, and $ \lambda = 
6/g_H^4 $. We 
do not show here the GFMC results of Hamer {\it et al.} \cite{hamer00}, 
which were derived from Creutz ratios on very small loops, and therefore 
subject to large finite-size effects. We see that the ELCE results are 
more accurate in the strong coupling region, but our PIMC results are 
better in the weak coupling region. The two sets of data do not agree 
below $ \sqrt{\lambda} = 2.1 $. This may be because our point at $ 
\sqrt{\lambda} = 2.00 $ (or $ \beta_E = 4.81 $) is too high, as noted 
above; but also, we do not expect precise 
agreement in this region, since the measurements correspond to different 
quantities. The PIMC estimates refer to the decay exponent of space-like 
Wilson loops, while the ELCE estimates are of the `axial' string tension, 
i.e. the energy per unit length of a `string' of flux along one axis. It 
is well known that these estimates may differ at strong coupling, where 
rotational invariance is broken; but at some intermediate coupling 
estimated \cite{kogut81} to be around $ \lambda = 5.4 $, a ``roughening'' 
transition 
\cite{hasenfratz81b,itzykson80,luscher81} takes place, after which 
rotational invariance is restored and different estimates of the string 
tension should coincide. Series expansions for the axial 
string tension cannot be continued past the roughening transition.

Again the crossover to weak coupling behavior occurs in the region 
$ \sqrt{\lambda} = 2.1 - 2.4 $, corresponding to $ \beta_E = 5.1 - 5.8 $. 
In the weak coupling region, the ELCE results are 
somewhat higher than our PIMC estimates, but are almost compatible within 
errors: we believe our present results are more reliable and accurate. 
They are also much more reliable than the GFMC results of Ref. \cite{hamer00}, 
mentioned above. 

We again see evidence for the approach to asymptotic scaling behavior in the region 
$ \sqrt{\lambda} > 2.3 $. The data give us a fit of
\begin{equation}
\sqrt{\sigma}  = (123 \pm 5)\Lambda_H .
\end{equation}
or
\begin{equation}
\sqrt{\sigma} = (1.34 \pm 0.05) \Lambda_{\mbox{\footnotesize mom}},
\end{equation}
using the conversion factors computed by Hasenfratz and Hasenfratz 
\cite{hasenfratz81}. This is much the same as our Euclidean estimate, and 
somewhat too high by comparison with Eq. (\ref{eq22}), presumably for the same 
reasons discussed above. It is considerably better than the Hamiltonian ELCE estimate $ \sqrt{\sigma} = (1.7 \pm 0.4) \Lambda_{\mbox{\footnotesize mom}} $ of Hamer {\it et al.} \cite{hamer86}, however.

\section{Conclusions}
\label{sec:conc}

We have demonstrated in this work that one can obtain reliable results for the Hamiltonian limit using the standard Path Integral Monte Carlo method for anisotropic lattices and extrapolating to $ \Delta \tau \rightarrow 0 $. We have calculated two quantities, the average (space-like) plaquette and the string tension. In both cases our results are a substantial improvement on other estimates calculated purely within the Hamiltonian formulation, although there is broad agreement between them. This demonstrates clear evidence of the universality between the Hamiltonian and Euclidean formulations of LGT.

In performing the extrapolation to the Hamiltonian limit we have found it very important to take into account the renormalization of the couplings $ \beta_\xi $ and $ \bar{\xi} $ in Eq. (\ref{ani_action_2}) at finite $ \xi $. The renormalization of $ \bar{\xi} $ has been discussed in some detail already by Klassen \cite{klassen98}. In our case this is only half the story, as there is also the renormalization of $ \beta_\xi $, which is related to the Euclidean coupling $ \beta_E $, through (\ref{karsch_betaxi}). As mentioned in the text, the discrepancy between the two couplings reaches a maximum around $ \xi = 3.6214 $, which is quite close to some of our most anisotropic data points, and has a large influence in terms of extrapolating our results to $ \xi \rightarrow \infty $. We expect that this renormalization will have a much smaller effect for improved actions \cite{morningstar97II,shakespeare98}, and hence utilizing such actions may be a neater way of removing the complication altogether. 

Our main motivation for examining anisotropic lattices was to investigate alternative Monte Carlo procedures for obtaining results in the Hamiltonian formulation, in view of the lack of a reliable quantum Monte Carlo algorithm for Hamiltonian LGT. We have found that the PIMC approach is superior to quantum Monte Carlo methods, in particular the GFMC algorithm. The GFMC algorithm was found previously \cite{hamer00} to have unacceptable systematic errors due to its dependence on a trial wave function. In this investigation we have found that the PIMC method gave more reliable results for the mean plaquette, and also gave good results for the string tension in the scaling regime, which the GFMC method \cite{hamer00} could not. 

The major disadvantage of PIMC is the necessity to make an extrapolation to the Hamiltonian limit $ \Delta \tau \rightarrow 0 $. This reduces the accuracy of the final results, and also is rather expensive in computer time, as several configuration sets must be generated to obtain one $ \beta_E $ point. With modern algorithms, however, the results are almost as accurate at large anisotropies as in this isotropic case, and the extrapolation can be made with confidence - see Figs. \ref{fig_avplaq_extrap} and \ref{fig_stringtens_extrap}. Thus we conclude that PIMC is the preferred Monte Carlo approach for estimating physical quantities in the Hamiltonian limit, just as in the Euclidean case. 

An obvious extension of the present work would be to attempt to estimate glueball masses using the PIMC approach, and to compare those with existing Hamiltonian results. We hope to attempt this in future work.

\appendix*
\section{Strong coupling expansion for SU(3) on anisotropic lattices}

A Wilson loop of size $ I \times J $ is defined by
\begin{eqnarray}
\nonumber
W(C) & = & \frac{1}{Z} \int [dU] \frac{1}{N} \chi \left( \prod_{i,j \in C} U_{ij} \right) \\
& & \times \prod_{P \in P_{ss}} e^{-S^{ss}(P) } \prod_{P \in P_{st}} e^{-S^{st}(P)} ,
\end{eqnarray}
where $ C $ denotes the contour of the Wilson loop, $ N $ is the dimension of the group matrices (in our case $ N = 3 $), $ U_{ij} $ denotes a group element lying between sites $ i $ and $ j $, $ Z $ is the normalizing factor, $ [dU] $ denotes the group integration over SU(3) matrices, $ \chi() $ takes character of its argument, the product over $ P_{ss} $ runs through all space-like plaquettes, while $ P_{st} $ denotes all time-like plaquettes. We also define
\begin{eqnarray}
S^{ss}(U) & = & - \frac{\beta_\xi}{\bar{\xi} N} \chi ( U) \\
S^{st}(U) & = & - \frac{\beta_\xi \bar{\xi}}{N} \chi ( U).
\end{eqnarray}
Using standard character expansion techniques \cite{creutz83}, we write
\begin{eqnarray}
e^{-S^{ss}(U)  } & = & \sum_{r=0}^\infty d_r \chi_r (U) \\
e^{-S^{st}(U)  } & = & \sum_{r=0}^\infty f_r \chi_r (U).  
\end{eqnarray}
The coefficients $ d_r $ and $ f_r $ may be found by inverting the relation
\begin{equation}
d_r = \int [dU] \chi^*_r (U) e^{-S^{ss} (U) },
\end{equation}
and similarly for $ f_r $. We only need the result for $ r = 0 $ and $ r= 3 $ however, which may be found in Eqs. (7) and (10) of Ref. \cite{smit82}. Defining
\begin{equation}
u(x) = \frac{1}{3}x + \frac{1}{6}x^2 - \frac{5}{72}x^4 - \frac{1}{24}x^5 + \frac{7}{720}x^6 + \dots, 
\end{equation}
then coefficients $ e_3 \equiv d_3/d_0 $ and $ g_3 \equiv f_3/f_0 $ are given by
\begin{eqnarray}
\label{e3def}
e_3 & = & 3u (\beta_\xi / 6 \bar{\xi} ) = \frac{\beta_\xi }{6 \bar{\xi}} + \frac{\beta^2}{72 \bar{\xi}^2 } + \dots \\
\label{g3def}
g_3 & = & 3 u (\beta_\xi \bar{\xi} / 6) = \frac{\beta_\xi \bar{\xi} }{6} + \frac{\beta^2 \bar{\xi}^2 }{72 } + \dots .
\end{eqnarray}
Performing expansions according to the diagrams described in Ref. \cite{creutz83}, we obtain for space-like Wilson loops
\begin{eqnarray}
\nonumber
W_{ss}(C) & = & \left( \frac{e_3}{N} \right)^{IJ} \Big[ 1 + 2 IJ  \left( \frac{g_3}{N} \right)^4 + 2 IJ  \left( \frac{e_3}{N} \right)^4 \\
\nonumber
& & + 2 IJ N \left( \frac{e_3}{N} \right)  \left( \frac{g_3}{N} \right)^4  + 2 IJ N  \left( \frac{e_3}{N} \right)^5 + \dots \Big] , \\
\label{Wssani}
\end{eqnarray}
while for time-like Wilson loops we obtain
\begin{eqnarray}
\nonumber
W_{st} (C) & = & \left( \frac{g_3}{N} \right)^{IJ} \Big[ 1 +  4 IJ \left( \frac{e_3}{N} \right)^2  \left( \frac{g_3}{N} \right)^2 \\
\nonumber
& & +   4 IJ N \left( \frac{e_3}{N} \right)^2  \left( \frac{g_3}{N} \right)^3 + \dots \Big] . \\
\end{eqnarray}
The string tension may then be computed using these results via $ \sigma a^2 = (1/IJ) \ln W(C) $. 

\begin{acknowledgments}
This work was supported by the Australian Research Council. We would like to thank the National Computing Facility for Lattice Gauge Theory for the use of the Orion supercomputer. 
DBL thanks Rudolf Fiebig for helpful discussions surrounding the
development and testing of anisotropic lattice gauge actions. TB thanks
A/Prof. Atsushi Hosaka for his hospitality while 
allowing access to the RCNP library in Osaka.

\end{acknowledgments}

\bibliography{paper}

\end{document}